\documentclass[aps,pra,twocolumn,notitlepage]{revtex4-1}

\usepackage{graphicx}

\begin{document}

\title{What are the observable effects of the physical properties of a quantum system?}

\author{Holger F. Hofmann}
\author{Taiki Nii}
\author{Masataka Iinuma}
\email{hofmann@hiroshima-u.ac.jp}
\affiliation{
Graduate School of Advanced Sciences of Matter, Hiroshima University,
Kagamiyama 1-3-1, Higashi Hiroshima 739-8530, Japan}

\begin{abstract}
In recent work (Nii et al., arXiv: 1603.06291; Iinuma et al., Phys. Rev. A 93, 032104 (2016)(arXiv: 1510.03958)) we have studied the relation between experimental outcomes and the physical properties represented by Hilbert space operators of a quantum system. We find that the values of physical properties are determined by the combination of initial and final conditions, which means that eigenstates and eigenvalues should not be misinterpreted as an exclusive set of possible realities. Here, we discuss the practical implications of these results and point out the importance of quantitative relations for a proper understanding of physical effects.
\end{abstract}


\maketitle

\section{Introduction}
Present theories of quantum measurements depend on the use of Hilbert space as a mathematical structure that is only loosely connected with experimental evidence by postulates regarding the unexplained physics of quantum state preparation and measurement outcomes. While these postulates provide rules for the practical application of quantum theory, they fail to explain how and why real physical objects can be described by such obscure rules of calculation. In the laboratory, the situation is opposite: it is only possible to observe quantum objects because the physics by which the object in question interacts with the instruments is known sufficiently well. Unfortunately, there is little effort to reconcile this practical knowledge about the physical object with its mathematical representation, leaving us with an inconsistent and paradoxical dualism of classical and quantum mechanical patterns of explanation. 

In general, the problems of quantum mechanics should be addressed by careful investigations of the relations between the observable properties of physical objects, since such relations describe the laws of causality by which we can know how quantum states are prepared or measured \cite{Hof14,Hof15,Hof16}. It is therefore not sufficient to interpret measurement outcomes only in terms of the statistics of a generally unknown input state. Instead, it is important to understand how both the initial state condition and the measurement outcome relate to the objective properties of the physical system represented by the Hilbert space formalism. 

As pointed out by Ozawa \cite{Oza03}, the Hilbert space formalism actually provides such a relation between the objective physical properties and the initial and final conditions of a quantum measurement in the form of the self-adjoint operators that represent the observable properties of the system. It is therefore possible to use Ozawa's definitions of measurement errors as a starting point for a more detailed investigation of the causality relations that govern the physics of quantum measurements. Here, we review and summarize our previous analysis of the role of causality and non-classical correlations in quantum measurements using the experimental and theoretical procedures of Ozawa's approach to measurement uncertainties \cite{Iin16,Nii16}. The results of this analysis may help us to understand why we should not identify eigenvalues with elementary objective realities, and where we need to modify and improve our understanding of the natural world in order to identify and appreciate the actual scientific content of quantum mechanics. 

\section{Quantification of effects}
A quantum measurement usually involves a rather complicated sequence of interactions between the actual quantum system and a number of auxiliary systems that constitute the technical implementation of the measurement. To avoid confusion, it is necessary to summarize the physical details of these interactions, focusing only on the relation between the objective physical properties in the input and the observable effect of these properties, which is finally registered as a measurement outcome $m$. However, it is impossible to identify the objective physical properties in the input using the terminology of quantum theory. Instead, the input state $\mid \psi \rangle$ provides a rather abstract mathematical description of the statistics of the input, without any clear reference to the various physical properties of the system. As a result, the input state defines the statistics of measurement outcomes without any indication of how the different outcomes relate to the quantitative properties of the system. Similarly, it is not necessarily clear how the measurement outcome $m$ relates to the complete set of physical properties of the system. The most microscopic characterization of the measurement is given by the measurement operator $\hat{E}_m$, which expresses the conditional probabilities of $m$ for any input state $\psi$ as
\begin{equation}
P(m|\psi) = \langle \psi \mid \hat{E}_m \mid \psi \rangle.
\end{equation} 
In a quantitative measurement of a physical property $\hat{A}$, it is necessary to associate a measured value $\tilde{A}_m$ with the outcome $m$, usually by attaching a scale to the readout. However, we need to make sure that the scale we are attaching to the outcomes $m$ really represents the quantity described by the observable $\hat{A}$. 

Significantly, the Hilbert space operator $\hat{A}$ represents the peculiar attachment of quantity to states, not just for the narrow case of eigenstates, but in all generality. The consequences are highly non-trivial, since we can now evaluate the magnitude of the quantitative difference between the objective property $\hat{A}$ and the measured value $\tilde{A}_m$ for any combination of input state and measurement outcome by using the average of the squared difference \cite{Oza03,Iin16,Nii16},
\begin{equation}
\label{eq:error}
\epsilon^2(A)= \langle \psi \mid (\tilde{A}_m-\hat{A}) \hat{E}_m (\tilde{A}_m-\hat{A}) \mid \psi \rangle.
\end{equation}
As we show in detail in \cite{Nii16}, the operator $\hat{A}$ cannot be replaced with a probability distribution over eigenvalue outcomes, since this would necessarily result in negative joint probabilities for the outcomes $m$ and the eigenvalues of $\hat{A}$. Instead, the error can be minimized by directly applying the quantitative measure of correlations between $m$ and $\hat{A}$ given by the operator product of Hilbert space \cite{Iin16}. These non-classical correlations minimize the error in Eq.(\ref{eq:error}) for measurement values of $\tilde{A}_m=A_{\mbox{opt.}}(m)$ with 
\begin{equation}
\label{eq:opt}
A_{\mbox{opt.}}(m) = \mbox{Re}\left[\frac{\langle \psi \mid \hat{E}_m \hat{A} \mid \psi \rangle}{\langle \psi \mid \hat{E}_m \mid \psi \rangle}\right]
\end{equation}
which is know to be the real part of the weak value of $\hat{A}$ for the combination of initial state $\psi$ and measurement outcome $m$ \cite{Hal04}.

Effectively, Eq.(\ref{eq:opt}) maximizes the non-classical correlation between the operator $\hat{A}$ with eigenstates $\mid a \rangle$ and the quantity $A_{\mbox{opt.}}(m)$ associated with the operators $\hat{E}_m$, which represent projections on completely different sets of eigenstates $\mid m \rangle$. These non-classical correlations define the proper quantitative relations between the physical properties of a quantum system and the measurement outcomes $m$.

\section{Deterministic relations between\\ outcomes and properties}
Of particular interest is the case of deterministic relations, where the quantitative errors $\epsilon^2(A)$ are exactly zero. In that case, the value of the physical quantity $\hat{A}$ is fully determined by the combination of initial condition $\psi$ and measurement outcome $m$. Oppositely, one can conclude that the measurement outcome $m$ has been caused by the combination of an initial quantity $B_\psi$ and the value $\tilde{A}_m=A_{\mbox{opt.}}(m)$ of the property $\hat{A}$. 

If the measurement in question is represented by a projection on an orthogonal set of states $\{ \mid m \rangle \}$, so that the measurement operators are given by $\hat{E}_m=\mid m \rangle \langle m \mid$, the condition for an error of zero can be written as
\begin{equation}
\langle m \mid (\tilde{A}_m - \hat{A}) \mid \psi \rangle = 0
\end{equation}
for all measurement outcomes $m$. This condition is automatically satisfied by $\tilde{A}_m=A_{\mbox{opt.}}(m)$ if the weak value of $\hat{A}$ is real,
\begin{equation}
\label{eq:Imzero}
\mbox{Im}\left[\frac{\langle m \mid  \hat{A} \mid \psi \rangle}{\langle m \mid \psi \rangle}\right] = 0.
\end{equation}
If the matrix representation of $\hat{A}$ in the basis $\{ \mid m \rangle \}$ has only real matrix elements, any state $\mid \psi \rangle$ with real coefficients $\langle m \mid \psi \rangle$ will satisfy this requirement. It is thus possible to find a large number of non-trivial cases where an error free value of $\hat{A}$ can be obtained from measurement outcomes $\mid m \rangle$ that are not eigenstates of $\hat{A}$, but relate to the value of $\hat{A}$ given by $\tilde{A}_m$ through a property defined by the initial state $\mid \psi \rangle$.

It is possible to separate the measurement information from the initial information mathematically by defining a pair of self-adjoint operators that represent the quantity $\hat{M}$ that determines the measurement outcome and the initial quantity $\hat{B}$ that defines the input state. The physical property $\hat{A}$ can then be given by the sum of the two quantities \cite{Nii16},
\begin{equation}
\label{eq:opsum}
\hat{A} = \hat{B} + \hat{M},
\end{equation}
where the operator $\hat{M}$ is given by the error free values $\tilde{A}_m$ and the eigenvalue $B_\psi$ as
\begin{equation}
\hat{M} = \sum_m (\tilde{A}_m - B_\psi) \mid m \rangle \langle m \mid.
\end{equation}
With this definition of physical properties, we find that the initial state $\mid \psi \rangle$ is an eigenstate of the difference between $\hat{A}$ and $\hat{M}$,
\begin{equation}
(\hat{A}-\hat{M})\mid \psi \rangle = B_\psi \mid \psi \rangle.
\end{equation}
In this sense, the initial condition $\psi$ determines the precise quantitative difference between the target observable $\hat{A}$ and the quantity $\hat{M}$ that determines the measurement outcome $m$. As a result, the quantum fluctuations of $\hat{A}$ in $\mid \psi \rangle$ are maximally correlated with the quantum fluctuations of $\hat{M}$, no matter which measurement procedure is used to evaluate the two quantities. 

\section{Physical properties as\\ elements of causality}
Measurement outcomes are the observable effects of the physical properties of the system. In the most general case, these effects cannot be traced back to error free values for the observables of the system. However, a maximally precise measurement, which is represented by an orthogonal set of state vectors in Hilbert space, provides exact relations between the initial state and all physical properties $\hat{A}$ that satisfy relation (\ref{eq:Imzero}). Quantum mechanics thus describes fully deterministic sum relations between a rather wide range of physical properties, even when these physical properties are represented by non-commuting operators. 

How do we know what the physical properties of a system are? Can we even know what a system is without first clarifying the essential properties of the system? When we talk about a ``particle,'' what are we really talking about? Perhaps we should dare to ask these questions - it may well be that they represent gaping holes in our understanding of physics that are only left undiscussed because we are too ashamed to openly confront our own ignorance. For instance, the word ``particle'' is itself a rather vague concept, implying that something sufficiently familiar to us can be reduced to elementary parts that nevertheless retain all of the properties associated with the more familiar object. In practice, we use the particle concept in quantum physics to describe the conserved quantitative causes of effects that appear to originate from a location - a center of mass or a center of charge. Under some circumstances, we can observe the change of location by observing a sequence of effects, and this results in the approximate assignment of trajectories. In the limit of state preparation and measurement, the best approximation of motion is a preparation of $\mid \psi \rangle$ at $t_0$ and a measurement of position $\hat{x}$ at $t_0+t$. If $\mid \psi \rangle$ is localized around $x=0$, it is natural to associate $\hat{x}/t$ with the velocity $\hat{v}$ of the particle. Indeed, it seems difficult to argue against this definition of velocity, given that the velocity is originally a measure of change, and only turns into a static property when we take the differential limit of $t \to 0$. 

It may be good to recognize that physical properties are not originally defined as static ``elements of reality,'' but are instead introduced as elements of transformations that describe the causality of dynamics \cite{Hof16}. In the present theoretical analysis, this fundamental aspect of physical properties is used to obtain an objective explanation of the causality relations between the initial and final effects of the system. In this sense, Eq.(\ref{eq:opsum}) means that the quantitative measurement outcome $\hat{M}$ is caused by the difference between the property $\hat{A}$ and the property $\hat{B}$ in the system. Essentially, the outcome $\hat{M}$ can be explained without any explicit reference to the physical property $\hat{M}$, since $\hat{M}$ is completely defined by its relation with $\hat{A}$ and $\hat{B}$. 

It is hard to see how a proper understanding of physics is possible if the relations between non-commuting physical properties are ignored. There is an infinite number of possible measurement scenarios, and most of them are described by non-commuting measurement operators. It is therefore important to recognize that the algebra of operators represents quantitative relations between observable effects that apply even when the operators representing initial states and final outcomes do not commute with each other. A truly scientific understanding of physical systems can only emerge from the identification of robust patterns that permit a unified explanation of all observable effects.

\vspace*{0.2cm}

\section{Conclusions}
Quantum mechanics has caused a lot of confusion with respect to the role and purpose of measurements in physics. Ideally, it should be possible to understand the physics of a system without any reference to a separate measurement theory, since the measurement merely maps the physical properties of the system onto the output. We therefore conclude that the object of interest should not be the measurement results themselves, but the physical properties of the system that caused these measurement results. Even in quantum mechanics, the physical explanation of the measurement outcomes must be related to the system through causality relations between the quantitative properties of the system and the measurement outcomes. 
In our analysis of quantum measurements, we have shown that the operator formalism of quantum mechanics provides such causality relations once it is understood that neither the eigenstates nor the outcomes of other measurements represent objective realities of the system \cite{Nii16}.


\end{document}